\documentclass[conference]{IEEEtran}
\pdfoutput=1 

\usepackage{amsmath,amssymb,amsfonts}
\usepackage{algorithmic}
\usepackage{graphicx}
\usepackage{textcomp}
\usepackage{xcolor}
\usepackage{pbalance}

\usepackage[utf8]{inputenc}
\usepackage[T1]{fontenc}

\usepackage{hyperref}
\hypersetup{hidelinks,unicode}
\usepackage{cleveref}
\usepackage[style=ieee, backend=biber, url=true, hyperref, maxnames=5, minnames=1, maxcitenames=5, mincitenames=1,defernumbers=true]{biblatex}

\makeatletter
\let\blx@rerun@biber\relax
\makeatother

\usepackage{acronym}
\usepackage{tabto}
\usepackage{enumerate}
\usepackage{siunitx}

\usepackage{tikz}
\usetikzlibrary{calc, arrows}
\usepackage{pgfplots}
\pgfplotsset{compat=1.18}

\usepackage[normalem]{ulem}
\useunder{\uline}{\ul}{}

\usepackage{orcidlink}

\DeclareBibliographyDriver{online}{%
  \usebibmacro{bibindex}%
  \usebibmacro{begentry}%
  \usebibmacro{author/editor+others/translator+others}
  \setunit{\labelnamepunct}\newblock
  \usebibmacro{title}%
  \newunit
  \printlist{language}%
  \newunit\newblock
  \usebibmacro{byauthor}%
  \newunit\newblock
  \usebibmacro{byeditor+others}%
  \newunit\newblock
  \printfield{version}%
  \newunit
  \printfield{note}%
  \newunit\newblock
  \printlist{organization}
  \newunit\newblock
  \usebibmacro{date}%
  \newunit\newblock
  \iftoggle{bbx:eprint}
    {\usebibmacro{eprint}}
    {}%
  \newunit\newblock
  \usebibmacro{url+urldate}%
  \newunit\newblock
  \usebibmacro{addendum+pubstate}%
  \setunit{\bibpagerefpunct}\newblock
  \usebibmacro{pageref}%
  \usebibmacro{finentry}}

\makeatletter
\renewcommand{\fnum@figure}{Figure~\thefigure}

\makeatother

\usepackage{xpatch}
\xpatchcmd\IEEEkeywords{---}{-}{}{}

\usepackage{ccicons}
\makeatletter
\def\ps@IEEEtitlepagestyle{
  \def\@oddfoot{\mycopyrightnotice}
  \def\@evenfoot{}
}
\def\mycopyrightnotice{
  {\footnotesize
    \begin{minipage}{0.8\textwidth}
    \centering
    Please cite as: \fullcite{selfref}.
    \end{minipage}
  }
}
\makeatother
\usepackage[shortcuts]{extdash} 

\DeclareBibliographyCategory{selfref}
\addtocategory{selfref}{selfref}

\crefname{subsection}{subsection}{subsections}
\Crefname{subsection}{Subsection}{Subsections}

\begin{document}

\title{\bfseries\Large Security Challenges for Cloud or Fog computing-Based AI Applications
}

\author{
\IEEEauthorblockN{~\\[-0.4ex]\large Amir Pakmehr\IEEEauthorrefmark{1}, Andreas A\ss muth\IEEEauthorrefmark{2}, Christoph P.\ Neumann\IEEEauthorrefmark{2}%
, and Gerald Pirkl\IEEEauthorrefmark{2}\\[0.3ex]\normalsize}

\IEEEauthorblockA{\IEEEauthorrefmark{1}%
Department of Computer and Information Technology Engineering\\
Qazvin Branch, Islamic Azad University, Qazvin, Iran\\
E-mail: {\tt amir.pakmehr@QIAU.ac.ir}}
\IEEEauthorblockA{\IEEEauthorrefmark{2}%
Department of Electrical Engineering, Media and Computer Science\\
Ostbayerische Technische Hochschule Amberg-Weiden, Amberg, Germany\\
E-mail: {\tt$\lbrace$a.assmuth\,|\,c.neumann\,|\,g.pirkl$\rbrace$@oth-aw.de}\\[1ex]}
}

\maketitle

\begin{abstract}
Security challenges for cloud or fog-based machine learning services pose several concerns. Securing the underlying cloud or fog services is essential, as successful attacks against these services, on which machine learning applications rely, can lead to significant impairments of these applications. Because the requirements for Artificial Intelligence applications can also be different, we differentiate according to whether they are used in the cloud or in a fog computing network. This then also results in different threats or attack possibilities. For cloud platforms, the responsibility for security can be divided between different parties. Security deficiencies at a lower level can have a direct impact on the higher level where user data is stored. While responsibilities are simpler for fog computing networks, by moving services to the edge of the network, we have to secure them against physical access to the devices. We conclude by outlining specific information security requirements for Artificial Intelligence applications.
\end{abstract}


\section{Introduction}

At the latest, since the presentation of ChatGPT by OpenAI in November 2022, the topic of Artificial Intelligence (AI) has been present and interesting even among a non-specialist audience. It is somewhat misunderstood that Machine Learning (ML) has already been used for more and more services in the private, commercial and industrial sectors in recent years~-- and the trend is rising. For many ML applications, cloud services are quite central as they provide a fast, scalable, flexible and cost-effective infrastructure for running sophisticated ML models and algorithms. Through them, enterprises can successfully and efficiently implement their ML projects. Some of the key benefits of cloud services for ML are:
\begin{enumerate}
    \item \textbf{Scalability}: Cloud services make it possible to scale up the required computing power to meet the requirements of the respective ML application. This is not only about the execution of the ML application, also the training of the models can be pushed by additional computing power or more available memory for larger amounts of data. Elasticity allows to  scale down the computing resources, when they are not needed any more.
    \item \textbf{Flexibility}: There is a great versatility in the ML services offered, like cloud-based machine learning development platforms in general, but also dedicated ML services for text-to-speech, speech-to-text, translation, conversations, automated image and video analysis, and many more. Cloud offerings include infrastructure, platform, and software, which can be customized to suit the needs of different users and applications. Different deployment options for ML applications exist, such as container orchestration, virtual machines, or serverless computing.
    \item \textbf{Cost efficiency}: Companies that deploy ML applications do not have to buy required hardware and perpetual licenses themselves or pay for its operation, but they can rent computing power or storage as well as subscription-based licenses. The Cloud Service Providers (CSP) offer more fine-grained cost models than traditional data centers. In combination with elasticity, they allow for pay-as-you-go or pay-as-you-grow approaches, which can result in lower costs. The CSPs offer their ML services worldwide, thus, international distribution of enterprise ML products becomes cost efficient.
    \item \textbf{Data management}: Cloud services can store and process large amounts of data that usually accompany modern ML applications.
    \item \textbf{Integration}: Cloud services, as a now established technology, offer a variety of other established tools and services, e.g., visualization tools, connection to databases, or workflow engines, making it easy to seamlessly integrate ML applications into existing web-based services or even an IT infrastructure.
\end{enumerate}
However, a generalization that AI and ML applications are only possible with cloud services is not permissible. There are also other areas of application for ML, like autonomous driving, in which a connection to cloud services is not continuously possible or does not make sense in large parts. In autonomous driving, the merging of image and radar data on the current traffic situation of a vehicle must take place in real time. Particularly when human life and limb are at stake, for example, when emergency braking is required, there is no time to first transfer all image and radar data to the cloud, analyzing it using ML algorithms, come up with the ``brake at once'' decision, and transmit the command to trigger the braking process to the vehicle. Thus, it is imperative to already have sufficient computing power and memory in the vehicle that all processing, computation and decision-making can take place on the spot. In so-called edge computing, the processing of data and the execution of applications is generally done on edge devices or devices close to the data sources, instead of being processed somewhere in the cloud. As the above example should make clear, criteria, such as real-time capability or minimal latency, are often crucial for such applications. Still, cloud services are part of the autonomous driving ecosystem, e.g., the higher-level control of traffic flow in a particular region, which uses swarm data retrieved from connected cars or the provision of map and navigation services in the vehicles.
\par 
An argument against cloud services and in favor of on-premise hardware commonly is better control over data protection and information security. However, even domains with strict regulations on security, like healthcare and banking, have adopted cloud services for some applications \autocite{sharma2021cloud}.
If regulations or trusted hardware considerations require in-depth control, private clouds involve setting up a cloud infrastructure that is dedicated to one's organization and is not shared with others.\par
Other reasons for having the processing of data and the execution of applications closer to the source of the data rather than in the cloud brings us to fog computing that aims to extend cloud computing capabilities to Internet of Things (IoT) devices and other edge devices, such as routers, switches, and gateways. It aims to provide a ``foggy'' layer of computing resources between the cloud and the edge, much like fog lies between the ground and the sky. The fog layer can help to reduce the latency and bandwidth requirements of cloud computing \autocite{Dastjerdi2016}.
\par
An example of an ML application deployed in the context of fog computing can be the processing of sensor data in a smart grid system. A smart grid is an electrical network that uses sensors and smart devices to monitor and optimize energy consumption. These sensors collect data, such as power consumption, network load, electricity price, power generation from renewable sources, and weather forecasts. In a typical fog computing architecture, the sensor data can be processed and analyzed at the edge devices in the smart grid. ML models trained on the sensor data can make predictions about energy demand and perform appropriate optimizations. By processing the data at the edge, real-time optimization of the power grid can be achieved without having to send all (possibly privacy-critical) data to a remote cloud. This can reduce latency and improve system efficiency. In addition, the fog computing network helps increase the security of the smart grid by eliminating the need to transmit data over public networks. Sensitive data remains on the edge device and can be better protected from potential attacks.\par 
An essential prerequisite for the correct functioning of ML applications is correctly working and reliable cloud services or fog computing networks. Conversely, it is clear that the compromise of cloud services or a fog computing network will lead to massive problems for ML applications. Therefore, in this paper, we would like to present the security challenges for ML applications that are based on cloud or fog computing and provide guidance and best practice recommendations on how to mitigate or control the respective threats.
\par 
The paper is structured in the following manner:
Section~\ref{sect-cloud-sec-challenges} discusses security challenges in cloud computing. Section~\ref{sect-fog-sec-challenges} presents security challenges in fog computing. Finally, Section~\ref{sect-ai-security-challenges} addresses special security challenges for ML applications in cloud or fog environments. The paper ends with a conclusion and an outlook on future work.

\section{Cloud computing Security Challenges}\label{sect-cloud-sec-challenges}
With regard to the correct functioning of ML services provided over the Internet, securing the underlying cloud services plays a very important role. Successful attacks against the cloud services on which ML applications rely can lead to significant impairments of these applications. But this is not the only reason why cloud services must be sensibly secured against cyberattacks. In practice, the fact that several parties are usually involved in the provision of cloud services often proves to be problematic. The classic Cloud Security Responsibility Model (CSRM) basically differentiates responsibility according to cloud vendor and user, distinguished for the service models Infrastructure as a Service (IaaS), Platform as a Service (PaaS) and Software as a Service (SaaS) \autocite{NSA2018}. Regardless of how many parties share responsibility for the cybersecurity of a cloud service, it is fundamental to ensure that responsibility at the interfaces between different parties in particular is clearly defined, because a security problem in one party's responsibility could potentially threaten the security of other parties' areas of responsibility. Due to the layer model of the system architecture, it is obvious that security deficiencies at a lower level have a direct impact on the higher level where, for example, user data is stored.\par
At the Cloud Computing 2019 conference, Süß et al. presented an overview of information security challenges for cloud services at the time and assessed them using the Common Vulnerability Scoring System (CVSS) \autocite{Suess2019}. However, when the Covid-19 pandemic began in early 2020, the security threats to cloud services changed. As many companies and organizations became more reliant on cloud services in a relatively short period of time because everyone's life was moved to the cloud, cybercriminals took advantage of this and tried to exploit vulnerabilities in cloud infrastructures. Very often, these were classic attacks that can also be used to attack other web services.

\subsection{Data Breaches}\label{subsec:breaches}
First, we look at attacks that specifically target data stored in the cloud. This could be customer data, secret company documents or medical records on patients. In a data breach, unauthorized access to sensitive data is given, which, of course, can have serious consequences for businesses and individuals \autocite{Barona2017}. Data breaches can be the result of unintentional exposure of sensitive data due to misconfigurations or weak security measures \autocite{Sabahi2011} or a targeted attack. Data breaches usually result in the confidence of customers and business partners in the affected company being shaken, often coupled with a serious loss of reputation among the general public. Since the occurrence of a data leak usually involves a violation of compliance requirements and laws, such as the EU General Data Protection Regulation (GDPR), the affected companies usually have to fear regulatory consequences. In addition to fines, there may also be indemnity claims with a simultaneous loss of revenue. A data breach often leads to further impairment of business activities. For example, it would be similarly bad if such corporate data were accidentally deleted by an inattentive employee or deliberately by an attacker.\par 
It should be noted that many cloud services that were used during the pandemic are still being used~-- which certainly makes sense. Although it is clear that by encrypting data that is stored in the cloud and implementation of a strict access control, one can guard against data breaches, their number and extent over the past two years~(see \autocite{Breaches2021} or \autocite{Breaches2022}, for example) is frightening, even though the first year of the Covid-19 pandemic has been called the ``worst year on record'' in terms of data leaks \autocite{Breaches2020}.

\subsection{Ransomware Attacks}
A so-called ransomware attack is an attack in which the attacker uses special malware to encrypt the victim's data and extort it by demanding payment for the surrender of the key. Probably the best known example of a ransomware attack was the WannaCry attack in 2017 which targeted computers running Microsoft Windows, encrypted the users data and demanded ransom payments in Bitcoin \autocite{wannacry2017}. While early ransomware variants often encrypted only a user's data stored on the local hard drive, variants soon developed that also encrypted connected disks or cloud-based storage \autocite{Watson2016}.
\par
In addition to this general description of ransomware, it should be noted that the ``Ransomware as a Service (RaaS)'' business model has gained significant importance in recent years. Here, ransomware developers sell their malware as a service to criminals. RaaS platforms often offer various options that allow criminals to create and execute their own ransomware campaigns \autocite{Kibet2022}. Such platforms often allow the customization of the ransomware, like the selection of targets or the determination of the ransom amount. Such services target less tech-savvy criminals, thereby enabling a wider range of people to run such ransomware campaigns. This increases the risk of ransomware attacks for all types of IT systems and, thus, also cloud services. We have already pointed out that in the case of cyberattacks on cloud services, responsibilities may be divided among several parties. Ransomware attacks are no exception in this regard. A user's data stored in the cloud could be encrypted as a result of catching a malicious ransomware on their PC or mobile device. In such a case, the responsibility is relatively clear and the CSP usually cannot help in such cases, unless the cloud storage service includes regular backups, and it is possible to restore a previous state of the data after removing the ransomware from the user's devices. In the event of an attack on the CSP during which the data of several (probably many) customers is encrypted, the responsibility lies with the CSP.\par 
The best protection against ransomware attacks is to keep all software up to date in combination with regular backups. It is mandatory to install security updates as soon as they become available, because adversaries often exploit vulnerabilities in outdated software. This also comprises anti-virus software being constantly updated. Backups are mandatory, because even if a user sees no other way out than to pay the demanded ransom, this is no guarantee to get their own data back intact, of course. How much do you trust the promise of a criminal who has blackmailed you?

\subsection{Distributed Denial of Service}
\begin{figure}[!h]
    \centering
    \pgfplotsset{scaled y ticks=false}
    \small
    \begin{tikzpicture}
	\begin{axis}[%
            height=4.5cm,
            width=\linewidth,
		ylabel={Number of Attacks},
		ytick={700000, 800000, 900000, 1000000},
		yticklabels={700k,800k,900k,1M},
		enlargelimits=0.15,
		legend style={at={(0.5,-0.2)}, anchor=north,legend columns=-1},
		ybar,
		symbolic x coords={Jan,Feb,Mar,Apr,May,Jun},
		xtick=data,
		]
		\addplot[draw=black!60, fill=black!15] coordinates {(Jan,714299) (Feb,678886) (Mar,811036) (Apr,751439) (May,748848) (Jun,690115)};
				
		\addplot[draw=black, fill=black!60] coordinates {(Jan,972552) (Feb,919865) (Mar,965642) (Apr,880998) (May,839539) (Jun,758349)};
						
		\legend{{1H 2020$\qquad$}, 1H 2021}
	\end{axis}
    \end{tikzpicture}
    \caption{Monthly DDoS attack frequency during the Covid-19 pandemic in 2020 and 2021 (Germany) according to \autocite{Netscout2021}.}
    \label{fig:ddos}
\end{figure}
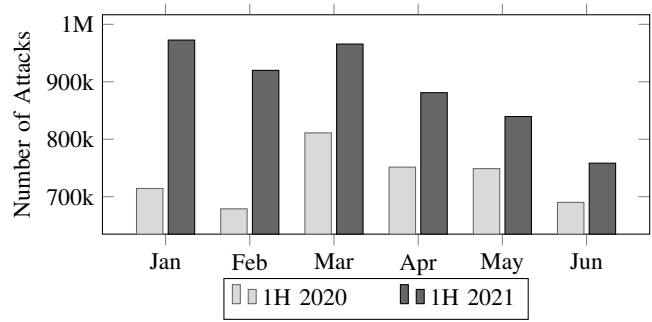

In a Distributed Denial of Service (DDoS) attack, a target system is flooded by mass requests. The bandwidth of the service's connection to the Internet is no longer sufficient, so that authorized users can no longer use the service in question. DDoS attacks are thus generally directed against the availability of services. During the Covid-19 pandemic, a very sharp increase in the number of DDoS attacks was observed. Figure~\ref{fig:ddos} illustrates this statement with a comparison of the numbers before or during the beginning (2020) and during the peak of the pandemic (2021) for the example of Germany. Cloud services are accessed via the Internet, which is why DDoS attacks that specifically target cloud services or their providers are a tried-and-tested means of preventing the use of these services, at least temporarily. 

The current trends, as described in \autocite{Cloudflare2023}, are that adversary often using cloud-based Virtual Private Servers (VPS) to set up botnets that are used to execute DDoS attacks. This setup is much more powerful than botnets based on vulnerable IoT devices, which we have seen in recent years. The attack durations decrease but, on the other hand, ransom DDoS attacks are on the rise. In such an attack, the adversaries extort ransom payments from the victim by threatening~-- for example, after a brief demonstration of their capabilities~-- to launch further or longer DDoS attacks if the victim does not pay the demanded sum.\par 
DDoS attacks may be prevented by using firewalls in combination with intrusion detection systems (IDS), traffic filtering and load balancing.

\subsection{Dependence on 3rd Party Software}\label{subsec:3rdparty}
The complexity of modern cloud services usually means that hardly any provider develops the software required for the service completely in-house. Instead, it is common to fall back on established and tested libraries, etc., which are developed and provided by 3rd parties. In case of open source software, it is possible to critically examine the source code of this integrated software and analyzing it in terms of vulnerabilities. Unfortunately, very often people trust that this 3rd party software has been thoroughly tested and is secure without checking this further. Vulnerabilities due to programming errors or deficiencies in the software design cannot be ruled out, however, and are often only discovered when the software has already been in use for a long time. A well-known example of a vulnerability in 3rd party software is Log4Shell, which was given a CVSS severity rating of 10.0, the highest score possible \autocite{Log4Shell2021}. Log4j is a Java-based logging utility which is used in open source and proprietary software alike and became a de facto standard for this purpose. The Log4Shell vulnerability allowed adversaries to remotely execute arbitrary code on the host system, e.g., to do some crypto mining on these systems. Quite a lot of services were affected, like Amazon Web Services (AWS) \autocite{AWS2021} or Apple's iCloud \autocite{Apple2021}, for example.\par
In principle, it does not matter which functionality is realized by 3rd party software. But in practice these are often security features whose technical characteristics are often cloud-specific, but are generally not conceptually new. To put this in concrete terms, the implementation of security roles, authorization rules, key or certificate management and methods in connection with Public Key Infrastructures (PKI) may be mentioned.\par 
The security of cloud services that use 3rd party software depends on the quality of such libraries, of course. For example, a vulnerability in an integrated authentication service can reveal personal data of customers of the CSP, which is a violation of the EU GDPR. In this example, it is initially irrelevant whether the data is generally accessible to anyone on the Internet (data breach) or whether getting access is much harder, but the data is stolen and published by an attacker. Responsible and liable in this case is the CSP, not the programmers of the (open source) library.\par 
The use of 3rd party software always poses a certain risk. To minimize this, it is therefore advisable to check any 3rd party software very carefully for vulnerabilities and also to test it extensively in interaction with one's own software components.

\subsection{Unsecured APIs}
APIs (Application Programming Interfaces) play an important role in the communication between cloud services and applications. If APIs are not sufficiently secured, they may become a potential security vulnerability. For example, an unsecured interface in a cloud API could result in confidential data being accessible to anyone (cf. data breaches, Subsection~\ref{subsec:breaches}). Secure APIs are in the interest of all parties involved, regardless of whether others access one's own code or data via these interfaces or whether we use libraries provided by others via such APIs (cf. Subsection~\ref{subsec:3rdparty}).\par 
A very good overview of security issues related to APIs is provided by the OWASP API Security Project \autocite{OWASPAPI2019}. They list the following security issues as their top 10:
\begin{enumerate}
    \item Broken Object Level Authorization
    \item Broken User Authentication
    \item Excessive Data Exposure
    \item Lack of Resources \& Rate Limiting
    \item Broken Function Level Authorization
    \item Mass Assignment
    \item Security Misconfiguration
    \item Injection
    \item Improper Assets Management
    \item Insufficient Logging \& Monitoring
\end{enumerate}
Several of these issues have already been mentioned in this paper and all of the items speak for themselves for readers who are familiar with information security. At this point, however, it should be noted that all these mentioned security problems are explained in detail in the report cited and appropriate countermeasures are proposed as well. At the time of writing, the OWASP API Security Project is working on the 2023 version of their top 10 list.

\subsection{Cloud-Native Security}
The Kubernetes documentation \autocite{kubernetesCloudNativeSecurity} summarizes cloud-native security as ``The 4C's of cloud Native security are cloud, Clusters, Containers, and Code''. The first and most famous container engine, Docker, optimizes for developer experience and ease of use and explicitly not for security. The Docker daemon requires root privileges and is a single executable monolith with a wide attack surface. This leads to exploits like DirtyCOW, however, it is hard to understand clearly the principle of its underlying vulnerability of Linux operating system, even for experienced kernel developers \autocite{wen2020analysis}.
\par
Alternative container engines are available now, e.g.\ Podman, OpenStack KataContainers, AWS Firecracker, or Google gVisor. Many of them focus on security and provide, i.e., a rootless mode. Lize Rice provides an introduction into applied container security \autocite{rice2020container}. Amazon introduced the Shared Responsibility Model \autocite{amazonSharedResponsibility}, which states that the provider is only responsible for security ‘of’ the cloud, while customers are responsible for security ‘in’ the cloud.
The \emph{10 Rules for Better Cloud Security} by GitGuardian \autocite{gitguardianRulesBetter} provide an entry point to measures that can be taken following the Shared Responsibility Model:
\begin{enumerate}
    \item Don’t overlook developer credentials (in public and private code repositories).
    \item Always review default configurations.
    \item List publicly accessible storage.
    \item Regularly audit access control.
    \item Leverage network constructs.
    \item Make logging and monitoring preventive.
    \item Enrich your asset inventory.
    \item Prevent domain hijacking.
    \item A disaster recovery plan is not optional!
    \item Limit manual configurations.
\end{enumerate}

Research about container security includes the creation of Trusted Execution Environments (TEE) for containers. Secure Linux containers can, e.g., be based on Intel SGX, as has been demonstrated by Arnautov et al.~\autocite{arnautov2016scone}.
\par
Finally, Kubernetes security is based on the 4Cs as mentioned above; all major CSPs provide guides to security and hardening of their Kubernetes environments \autocite{kubernetesCloudNativeSecurity}. Areas of concern for workload security in Kubernetes are, e.g., role-based access control (RBAC) authorization, application secrets management and encrypting them at rest, ensuring that pods meet defined pod security standards \autocite{gitguardianKubernetesSecurity}, and network policies.

\section{Fog computing Security Challenges}\label{sect-fog-sec-challenges}
Fog computing, a term coined by Cisco~\autocite{Cisco2015wp}, is a distributed computing paradigm that bridges the gap between cloud computing and IoT devices. Rather than pushing all data to a remote cloud for processing, in fog computing, computations are instead carried out closer to the source of data~-- on the IoT devices themselves or on local edge servers. This minimizes the latency involved in long-distance data transport, optimizes system efficiency and improves real-time capabilities. By bringing computation and storage closer to the data sources, fog computing addresses issues like bandwidth constraints, latency, and security concerns that can be associated with cloud computing~\autocite{bonomi2012}. This approach is particularly beneficial for high mobility technologies like the IoT and Vehicular Ad-hoc Networks (VANETs), as it provides faster communication and software services to users. By reducing the distance between devices and computing resources, fog computing offers lower latency and improved quality of service compared to traditional cloud computing \autocite{Dastjerdi2016}\autocite{Mahmud2017}.
\par 
While fog computing shares some characteristics with cloud computing, it differs in several ways, such as balancing central and local computing, storage, and network management. This balance allows fog computing to offer more efficient, real-time control and improvements for various systems, including healthcare, traffic patterns, parking
systems, and more. But, of course, fog computing is not without its challenges. Some of its limitations include lower resources compared to cloud computing, higher latency in certain cases, energy consumption concerns, load balancing, data management, and security threats \autocite{Cisco2015}.
\par 
Based on the afore mentioned characteristics, fog computing may be seen as an addition to traditional cloud computing. And in the context of this paper, these characteristics allow choosing between cloud or fog computing as a basis for specific ML applications or projects.\par 
There are several security threats to fog computing that are comparable or similar to those to cloud computing. In general, fog computing involves a distributed network of devices, which increases the risk of network disruptions and downtime. So, besides data confidentiality, authenticity, and integrity, ensuring high availability is crucial to guarantee uninterrupted service delivery. Attacks can hinder the proper functioning of fog computing systems and may lead to unauthorized access, data leakage, or system failures \autocite{Khan2017}. To mitigate these attacks, fog computing systems must implement robust security measures, such as strong encryption, intrusion detection and prevention, access control, and continuous monitoring. Additionally, ensuring compliance with security standards and best practices can help minimize the risk of security breaches in fog computing environments \autocite{Aljumah2018}. A couple of security threats to fog computing have already been described in Section~\ref{sect-cloud-sec-challenges}. For example, DDoS attacks against fog computing networks aim to overwhelm fog nodes or networks with excessive traffic, causing disruptions and impacting services \autocite{Mukherjee2017}. Additionally, there are other classical network attacks, like Man in the Middle (MITM) or replay attacks. A MITM attack in fog computing involves intercepting and manipulating communications between legitimate components, compromising the system's integrity, confidentiality, and availability \autocite{Aliyu2018}. A replay attack is a type of security threat where an adversary captures and retransmits previously exchanged messages between parties in a communication session, making it seem as if they are the legitimate sender \autocite{Hosseinzadeh2019}. In the context of fog computing, an adversary may impersonate end devices or the fog broker to carry out this attack. During a replay attack, the adversary neither needs to understand the content of the captured messages nor decrypt any encrypted data; they simply replay the messages to exploit the system. This could lead to various negative consequences, such as unauthorized access, data manipulation, or disruption of services.\par 
In the following, we focus on threats and attacks that are more specific to fog than to cloud computing.

\subsection{Physical Attacks}
A physical attack in fog or edge computing \autocite{Alwakeel2021} involves compromising the physical hardware of the system, such as servers or other devices. This can be particularly problematic in these systems because their infrastructure is distributed across various geographical locations. If the physical protection of these devices is inadequate, it could allow for tampering or damage. Since each device or server typically serves a local geographical area, any physical attack can disrupt services within that specific area. Hence, it's crucial to implement strong physical security measures alongside cybersecurity measures in fog computing.

\subsection{Fog and User Impersonation Attack}
This is a type of cyberattack where an adversary poses as another device or user on a network in order to launch attacks against network hosts, steal data, spread malware, or bypass access controls. This is a particularly insidious type of attack because it can be very difficult to detect, as the adversary is using credentials that are considered valid within the system. Impersonation attacks in fog computing can disrupt the communication between fog nodes and end devices, leading to miscommunication, data theft, or even disruption of service. As a countermeasure, Tu et al.\ suggest combining physical layer security techniques with a reinforcement learning algorithm to improve the security against impersonation attacks and optimize the decision-making process for distinguishing between legitimate and unauthorized entities \autocite{Tu2018}.

\subsection{Malicious Fog Nodes Attacks}
A malicious fog node can compromise network operations through various attacks, affecting the reliability of fog-to-fog collaborations. Also, identifying malicious fog devices in fog computing is crucial. To prevent malicious fog node issues, organizations should implement a comprehensive security approach including authentication and authorization, data encryption, secure communication protocols, intrusion detection systems, trust management, regular monitoring, network segmentation, access control, and an incident response plan. These strategies help reduce risks, enhance overall security, and maintain system resilience. Consequently, achieving comprehensive protection against attacks becomes challenging, as it involves granting limited privileges and processing data. Finding appropriate countermeasures is the subject of current research, as examples we refer to Al-Khafajiy et al.~\autocite{Alkhafajiy2020} and Ke Gu et al.~\autocite{Gu2022}. The latter present a fog computing-based VANET, in which a scheme is used to detect malicious nodes (vehicles or devices with harmful intent). In their approach, the fog server computes a reputation score for each potentially harmful node. This score is determined by examining the relationship between the data collected from the node and the overall network structure. By analyzing these factors, the fog server can more accurately identify and flag nodes that may pose a threat to the network's security and performance.

\subsection{Rogue Fog Nodes}
A rogue fog node attack is when a malicious node pretends to be a legitimate fog node and joins the network to perform attacks, such as eavesdropping, data theft, or denial of service~\autocite{Yi2015}.
\par 
To prevent rogue fog node attacks, the following measures can be taken:
\begin{itemize}
    \item Authentication and authorization: Fog nodes should be authenticated and authorized before they are allowed to join the network. This can be achieved by implementing secure boot and mutual authentication mechanisms.
    \item Encryption: Sensitive data transmitted between fog nodes should be encrypted to prevent eavesdropping and data theft.
    \item Trust Management: Trust management protocols can be used to evaluate the trustworthiness of fog nodes. This can be based on the node's behavior, reputation, and credentials.
    \item Network Segmentation: Segmentation of the network can be used to isolate the fog nodes that are vulnerable to attacks. This can help in containing the attack and minimizing the damage.
    \item Continuous Monitoring: Continuous monitoring of the network can be used to detect any unauthorized fog node that joins the network. This can be achieved by monitoring network traffic, node behavior, and system logs.
\end{itemize}

\subsection{Ephemeral Secret Leakage Attack}
In the realm of fog computing, the Ephemeral Secret Leakage Attack~\autocite{Dewanta2020} presents a notable risk due to the distributed architecture and sensitive data often involved. This attack, based on the Canetti-Krawczyk adversary model~\autocite{Canetti2001}, assumes that an adversary can access one of the secret keys (short-term or long-term) used for secure communication between devices. If an adversary reveals a session key (a temporary encryption key), they can decipher all data exchanged during that session, leading to a potential security breach. Therefore, implementing robust cryptographic protocols and effective key management strategies is crucial for maintaining security in fog computing systems.

\section{Special Security Challenges for AI Applications}\label{sect-ai-security-challenges}

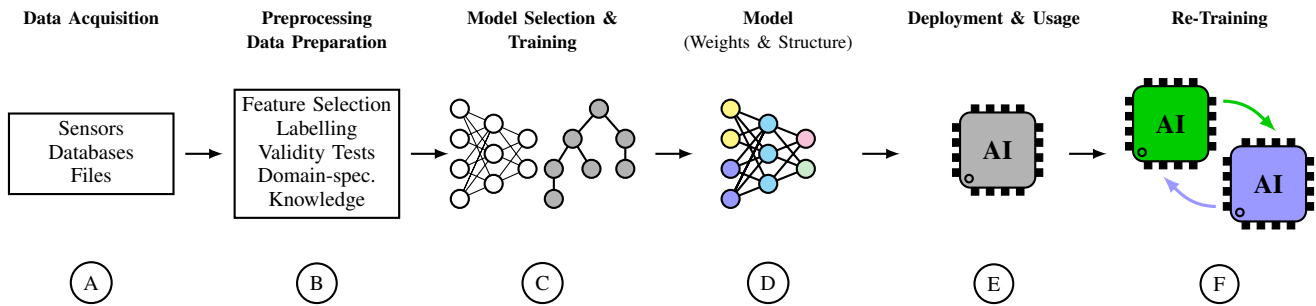
\begin{figure*}[!hbt]
    \centering
    \begin{tikzpicture}

        \newcommand{\chip}[3]{
            \foreach \no in {1,3,5,7}{%
                \draw[fill=black] (#1-1mm, #2+0.5mm+1mm*\no) rectangle (#1+1.1cm, #2+1.5mm+1mm*\no);
                \draw[fill=black] (#1+0.5mm+1mm*\no, #2-1mm) rectangle (#1+1.5mm+1mm*\no, #2+1.1cm);
            };
            \draw[thick, fill=#3, rounded corners] (#1, #2) rectangle (#1+1cm, #2+1cm);
            \draw[thick] (#1+1.2mm, #2+1.2mm) circle (0.5mm);
            \node at (#1+0.5cm, #2+0.5cm) {\bfseries AI};
        }
        \begin{scope} 
            \footnotesize
            \node[anchor=north] at (1.25, 4) {\bfseries\scriptsize Data Acquisition};
            \node[text width=2cm, text centered, draw, thick] at (1.25, 2) {Sensors\vphantom{j}\\Databases\vphantom{j}\\Files\vphantom{j}};
            \node[anchor=south, draw, thick, circle] at (1.25, 0) {A};
        \end{scope}
        \begin{scope} 
            \footnotesize
            \node[anchor=north, text width=2.25cm, text centered] at (4.25, 4) {\bfseries\scriptsize Preprocessing\\Data Preparation};
            \node[text width=2cm, text centered, draw, thick] at (4.25, 2) {Feature Selection\vphantom{j}\\Labelling\\Validity Tests\\Domain-spec. Knowledge};
            \node[anchor=south, draw, thick, circle] at (4.25, 0) {B};
        \end{scope}
        \begin{scope} 
            \footnotesize
            \node[anchor=north, text width=2.25cm, text centered] at (7.25, 4) {\bfseries\scriptsize Model Selection \&\\Training};
            \draw (6.15, 2.6) -- (7.05, 2.2) (6.15, 2.6) -- (6.6, 2) (6.15, 2.6) -- (6.6, 1.6) (6.6, 2.4) -- (6.15, 2.2)  (6.6, 2.4) -- (6.15, 1.8) (6.6, 2.4) -- (6.15, 1.4) (6.6, 2.4) -- (7.05, 1.8) (7.05, 2.2) -- (6.6, 2) (7.05, 2.2) -- (6.6, 1.6) (6.15, 2.2) -- (7.05, 1.8) (6.15, 1.8) -- (6.6, 2) (6.15, 1.8) -- (6.6, 1.6) (6.15, 1.4) -- (7.05, 1.8) (6.15, 1.4) -- (6.6, 2) (6.15, 1.4) -- (6.6, 2.4);
            \foreach \xpos/\ypos in {6.15/1.4, 6.15/1.8, 6.15/2.2, 6.15/2.6, 6.6/1.6, 6.6/2, 6.6/2.4, 7.05/1.8, 7.05/2.2}{%
                \draw[thick, fill=white] (\xpos, \ypos) circle (1.2mm);
            };
            \draw[thick] (8, 2.6) -- (8.35, 2.2) -- (8.35, 1.8) (8, 2.6) -- (7.65, 2.2) -- (7.4, 1.8) -- (7.4, 1.4) (7.65, 2.2) -- (7.9, 1.8);
            \foreach \xpos/\ypos in {8/2.6, 8.35/2.2, 7.65/2.2, 8.35/1.8, 7.4/1.8, 7.9/1.8, 7.4/1.4}{%
                \draw[thick, fill=black!30] (\xpos, \ypos) circle (1.2mm);
            };
            \node[anchor=south, draw, thick, circle] at (7.25, 0) {C};
        \end{scope}
        \begin{scope} 
            \footnotesize
            \node[anchor=north, text width=2.5cm, text centered] at (10.25, 4) {\scriptsize\textbf{Model}\vphantom{j}\\(Weights \& Structure)};
            \draw[thick] (9.75, 2.6) -- (10.75, 2.2) (9.75, 2.6) -- (10.25, 2) (9.75, 2.6) -- (10.25, 1.6) (10.25, 2.4) -- (9.75, 2.2)  (10.25, 2.4) -- (9.75, 1.8) (10.25, 2.4) -- (9.75, 1.4) (10.25, 2.4) -- (10.75, 1.8) (10.75, 2.2) -- (10.25, 2) (10.75, 2.2) -- (10.25, 1.6) (9.75, 2.2) -- (10.75, 1.8) (9.75, 1.8) -- (10.25, 2) (9.75, 1.8) -- (10.25, 1.6) (9.75, 1.4) -- (10.75, 1.8) (9.75, 1.4) -- (10.25, 2) (9.75, 1.4) -- (10.25, 2.4);
            \foreach \xpos/\ypos/\col in {9.75/1.4/blue!40, 9.75/1.8/blue!40, 9.75/2.2/yellow!60, 9.75/2.6/yellow!60, 10.25/1.6/cyan!40, 10.25/2/cyan!40, 10.25/2.4/cyan!40, 10.75/1.8/green!60!black!20!white, 10.75/2.2/magenta!30}{%
                \draw[thick, fill=\col] (\xpos, \ypos) circle (1.2mm);
            };
            \node[anchor=south, draw, thick, circle] at (10.25, 0) {D};
        \end{scope}
        \begin{scope} 
            \footnotesize
            \node[anchor=north, text width=2.5cm, text centered] at (13.25, 4) {\bfseries\scriptsize Deployment \& Usage};
            \normalsize
            \chip{12.8cm}{1.55cm}{black!30}
            \footnotesize
            \node[anchor=south, draw, thick, circle] at (13.25, 0) {E};
        \end{scope}
        \begin{scope} 
            \footnotesize
            \node[anchor=north] at (16.25, 4) {\bfseries\scriptsize Re-Training};
            \normalsize
            \chip{15.1cm}{1.9cm}{green!80!black}
            \chip{16.4cm}{1.1cm}{blue!40}
            \footnotesize
            \begin{scope}[fill=green!80!black, draw=green!80!black]
                \draw[very thick, -latex] (16.3, 2.7) to[out=0, in=135] (17, 2.3);
            \end{scope}
            \begin{scope}[fill=blue!40, draw=blue!40]
                \draw[very thick, -latex] (16.2, 1.3) to[out=180, in=315] (15.5, 1.7);
            \end{scope}
            \node[anchor=south, draw, thick, circle] at (16.25, 0) {F};
        \end{scope}
        \foreach \xpos in {2.5, 5.5, 8.75, 11.5, 14.25}{%
            \draw[thick, -latex] (\xpos, 2) -- ++(0.5, 0);
        };
    \end{tikzpicture}
    \caption{Typical AI workflow in different steps (A to F). Note: The deployment and application steps typically run on different systems.}
    \label{fig:aiworkflow}
\end{figure*}
\subsection{Data Representing ML models}
Some of the previously mentioned attacks targeted data stored in the cloud or in a fog computing network. In addition to the consideration that this data is, for example, personal data of customers, which is then processed by the ML application, there is another relevant aspect. A crucial prerequisite for successful ML projects is very often a sufficiently large amount of training data. ML models are only as good as the quality of the training data. In many areas where ML methods have not yet been applied or in the case of new business models, no training data are available at the beginning. These often have to be created laboriously at first, which on the one hand may mean a large number of measurements to generate a sufficiently large sample set and on the other hand often means the manual labeling of the training data. Against the background of the threats and attacks discussed earlier, it should therefore be emphasized that ML models and their training data are very valuable assets. Unavailable models or training data due to a DDoS attack can lead to severe business interruptions. But even worse would be if models or training data that are exposed on the Internet or stolen fall into the hands of a competitor. This could even spell the end for a company whose business model is based on such ML projects.

\subsection{Special AI-Related Security Issues}

The special situation arising when working with AI algorithms is the way the models are trained, deployed, integrated and used in industrial environments (cf. Figure~\ref{fig:aiworkflow}).\par

Typically, the models are centrally trained on special high performance computers or servers and after training and evaluation transferred to the application server. Referring to Figure~\ref{fig:aiworkflow}, Step~A, the data scientist and co-workers create the data set extracted from typical information sources such as sensors, databases and image archives. In Step~B, the data set is checked for validity, activities such as annotation (class assignment), feature extraction and the integration of domain-specific additional knowledge expand the data set in this step. Next, various models and AI architectures are applied to the data set and are then evaluated in Step~C. Usually, the approach/model with the highest robustness and accuracy is used and deployed. Depending on the data set, the time needed to train a model can vary between minutes or several days on high performance computers. Step~D addresses the fully trained model. The parameters of the model represent possible clusters or class memberships (the \textit{intelligence}). Step~E covers the deployment and application of the model. This includes the data preparation steps in the production environment, feature calculation and forward propagation of the feature vector in the AI model. The application runs on an application server, which usually requires a different level of security. A detached Step~F is the re-training of the AI model: new aspects that have arisen either through extension of the use case or through additional data during operation must be integrated into the model. Usually, not the entire system is re-trained, but parts of the upper layers of a network are algorithmically adapted.\par
We now consider the workflow shown in Figure 2 against the background of an industrial application, for example in a modern production line. Concerning the security of the AI model, three different scenarios can occur. In the \textbf{poisoned data set} scenario, the adversary, e.g. a malicious insider, has inserted harmful information into the data set that does not match the desired class and thus negatively influences and disrupts the AI structure after the training process. In general, ML poisoning attacks refer to the manipulation of data used for the (re-)training of ML models, e.g., in a fog environment~\autocite{Qi2021}. Especially in edge systems, basic AI training is performed at a central processing system due to the lack of processing resources. The generalized model is then deployed on the edge system and adapted to the application requirements using smaller local data sets or calibration steps. This local adaption process is prone to attacks as the training data is locally gathered without any supervision by experts.\par 
To prevent ML poisoning attacks, here are some countermeasures that can be taken:
\begin{itemize}
    \item \textbf{Use of Secure Data Sources}: Data sources must be secure and access to them should be restricted to authorized personnel only. It must be possible to check and verify the validity of the data at any time.
    \item \textbf{Data Sanitization}: The data should be checked and cleaned before it is used to train ML models. Any data that is found to be suspicious or anomalous should be removed. It may not be possible to automate this, but must be done manually.
    \item \textbf{Anomaly Detection}: Anomaly detection techniques can be used to detect any malicious data in the training data set. AI-based anomaly detection based on autoencoders, recurrent neural networks (RNN) or generative adversarial networks (GAN) can be considered as state of the art.
    \item \textbf{Ensemble Learning}: Ensemble learning is a technique where multiple ML models are trained on different subsets of the data. This can make it harder for adversaries to manipulate the data in order to affect the overall prediction.
    \item \textbf{Continuous Monitoring}: Continuous monitoring of the ML model's behavior is essential to detect any unusual or unexpected outcomes. Any anomalies should be investigated and addressed promptly.
\end{itemize}
In the \textbf{compromised AI model} scenario, the adversary changes the trained weights (parameters) of the AI network which leads to falsified outputs (cf. Step~D). Therefore, security measures to ensure the integrity of the data must be used in order to prevent manipulation. Of course, these measures must not interfere with the re-training of the model (cf. Step~F). It is conceivable to switch off these measures during re-training, but this requires that re-training is thoroughly monitored and secured against unauthorized access. If this is not possible or does not make sense, e.g., because the re-training is automated, a trust management system should be considered that evaluates the trustworthiness of the data for re-training and detects manipulated model parameters.\par 
The third scenario addresses deployment, integration and utilization of the AI system in the production environment (cf. Step~E). Information \textbf{inputs and results of the AI network} might as well be compromised: Either false/noisy information is presented to the network (input, e.g. by manipulated sensors) or the results are falsified and thus passed on incorrectly. In both cases, this interferes with the production steps that follow. Statistical analysis of the production can detect these kind of attacks.\par 
The above-mentioned security precautions can be introduced at various points in the processing architecture. In a fog environment, for instance, edge systems act as supervisors for local information sources; status information forwarded from edge nodes to central units must be checked and validated before integration into central data sets.\par 
In addition to the organizational challenges in the use of AI algorithms, there are also semantic problems: If habits change in the application field, this leads to sudden changes that trigger anomaly detection. Low-threshold changes, such as those applied by adversaries in the network area, may undermine the anomaly detection process. It is therefore necessary to weigh up the sensitivity of such approaches.

\subsection{Special Attacks on Language Models}
In this subsection, we focus on special attacks on language models.\par 
Suppose an attacker has access to multiple snapshots of an ML model, such as predictive keyboards. Then these snapshots can reveal detailed information about the change in training data used to update the model. This is called a model update attack. Zanella-Béguelin et al. analyzed information leakage in practical applications where language models are frequently updated, for example, by adding new data, deleting user data to meet privacy requirements, or matching private data with that of public, pre-trained language models. They developed two new metrics to analyze the information leakage, which now enables them to perform this kind of leakage analysis unsupervised \autocite{zanella2020}.
\par 
Tab attacks are attacks on language models that rely on autocompletion and in which the adversary attempts to cause the model to provide unwanted suggestions or results. So, these attacks target text recognition systems or try to figure out the robustness of language models. This involves an attempt to intentionally deceive the language model by deliberately inserting false or misleading information or creating distortions in the input data. Large language models are capable of memorizing rare training samples,
which poses serious privacy threats in case the model is trained on confidential user content. Inan et al.\ have developed a methodology for checking a language model for training data leaks. This enables the creator of the model to determine to what extent training examples can be extracted from the model in a practical attack. And the owner of the model is able to verify that deployed countermeasures work as expected, and thus that their model can be used securely \autocite{Inan2021}.

\section{Conclusion and Future Work}
In this paper, we have illustrated the dependency of AI applications on underlying cloud or fog-based services. Attacks against the cloud services or fog computing networks on which current AI applications are built will inevitably result in difficulties, data breaches, failures, or malfunctioning of the AI applications. AI is one of the current hot topics, resulting in high demand for related services. This makes them an attractive target for cybercriminals: they can try to prevent access to AI services on the Internet in order to extort a ransom from the service provider; they can also try to steal training data or complete ML models in order to have the owners pay for getting their data back or sell them to others, e.g., to competitors, at the highest bid.\par 
The interplay between AI and information security promises huge potential for future applications and research. For example, this paper did not even address how AI methods could also be used in order to support threat analysis of systems or penetration testing. It is already possible to use language models to generate phishing emails optimized for a specific target. Due to this huge potential of the interaction of AI and information security, we intend to continue to be active in these areas in the future.

\renewcommand*{\bibfont}{\footnotesize}
\setlength{\labelnumberwidth}{0.45cm}
\printbibliography[notcategory=selfref]

\end{document}